\documentclass[conference]{IEEEtran}
\IEEEoverridecommandlockouts

\makeatletter
\newcommand{\linebreakand}{%
  \end{@IEEEauthorhalign}
  \hfill\mbox{}\par
  \mbox{}\hfill\begin{@IEEEauthorhalign}
}
\makeatother

\usepackage{cite}
\usepackage{amsmath,amssymb,amsfonts}
\usepackage{algorithmic}
\usepackage{algorithm}
\usepackage{graphicx}
\usepackage{textcomp}
\usepackage{comment}
\usepackage{svg}
\def\BibTeX{{\rm B\kern-.05em{\sc i\kern-.025em b}\kern-.08em
    T\kern-.1667em\lower.7ex\hbox{E}\kern-.125emX}}
\begin{document}

\title{Utilizing Dynamic Time Warping for Pandemic Surveillance: Understanding the Relationship between Google Trends Network Metrics and COVID-19 Incidences}

\author{
\IEEEauthorblockN{Michael T. Lopez II}
\IEEEauthorblockA{\textit{Ateneo Social Computing Science Laboratory} \\
\textit{Department of Information Systems and Computer Science} \\
\textit{Ateneo de Manila University}\\
Quezon City, Philippines \\
michael.lopez@student.ateneo.edu}
\and
\IEEEauthorblockN{Cheska Elise Hung}
\IEEEauthorblockA{\textit{Ateneo Social Computing Science Laboratory} \\
\textit{Department of Information Systems and Computer Science} \\
\textit{Ateneo de Manila University}\\
Quezon City, Philippines \\
cheska.hung@student.ateneo.edu}
\and

\linebreakand
\IEEEauthorblockN{Maria Regina Justina E. Estuar}
\IEEEauthorblockA{\textit{Ateneo Social Computing Science Laboratory} \\
\textit{Department of Information Systems and Computer Science} \\
\textit{Ateneo de Manila University}\\
Quezon City, Philippines \\
restuar@ateneo.edu}
}

\maketitle

\begin{abstract}
The premise of network statistics derived from Google Trends data to foresee COVID-19 disease progression is gaining momentum in infodemiology. This approach was applied in Metro Manila, National Capital Region, Philippines. Through dynamic time warping (DTW), the temporal alignment was quantified between network metrics and COVID-19 case trajectories, and systematically explored 320 parameter configurations including two network metrics (network density and clustering coefficient), two data preprocessing methods (Rescaling Daily Data and MSV), multiple thresholds, two correlation window sizes, and Sakoe-Chiba band constraints. Results from the Kruskal-Wallis tests revealed that five of the six parameters significantly influenced alignment quality, with the disease comparison type (active cases vs. confirmed cases) demonstrating the strongest effect. The optimal configuration, which is using the network density statistic with a Rescaling Daily Data transformation, a threshold of 0.8, a 15-day window, and a 50-day radius constraint, achieved a DTW score of 36.30. This indicated substantial temporal alignment with the COVID-19 confirmed cases data. The discoveries demonstrate that network metrics rooted from online search behavior can serve as complementary indicators for epidemic surveillance in urban locations like Metro Manila. This strategy leverages the Philippines' extensive online usage during the pandemic to provide potentially valuable early signals of disease spread, and offers a supplementary tool for public health monitoring in resource-limited situations.
\end{abstract}


\section{Introduction}
The Philippines received long-term significant damage from the coronavirus disease (COVID-19) when it was one of the first nations in Southeast Asia that got infected in February 2020 \cite{amitEarly2021}. It is worth looking that those in the medical front lines against the disease in April 2020 were the most affected \cite{deveroPrevention2021}. The early phase of the pandemic logged 8,212 COVID-19 cases in the entire country, wherein about 24.7\% of them were healthcare workers and 35 of those in the medical field perished \cite{leyvaPandemic2024}. At the end of 2020, the Philippines had 451,839 confirmed cases and 8,812 deaths, which makes the case-fatality rate to be 1.95\% \cite{worldhealthorganizationPhilippines2020}. Through a heightened sense of urgency, the national government ushered more ramped up activities in contact tracing and testing capacity \cite{worldhealthorganizationPhilippines2020}. It seems this was a positive trait until the Delta variant led to a new wave of cases for the new year that caused the mortality rate to be 33\% above normal by the end of 2021 \cite{migrinoUsing2023}.

This pandemic also caused a deep recession in the Philippine economy. The gross domestic product in 2020 shrank by 9.5\% due to the strict community lockdowns, spearheading businesses to inactivity \cite{altercontactsLockdown}. The sharp economic decline, driven by suppressed consumption and investment, was among the worst in the Southeast Asian region for that year \cite{amitEarly2021}. One of those contributing factors were one of the most stringent lockdowns worldwide. The government had its initial classification called the ``Enhanced Community Quarantine'' imposed on the National Capital Region (also known as Metro Manila), wherein thereafter expanded to the entire Luzon island \cite{amitEarly2021}. Under these circumstances, residents were ordered to stay at home, public transport and non-essential businesses were shut, mass gatherings were banned, and the police and military forces administered the curfews nationwide \cite{deveroPrevention2021}.


The Philippines also had an extensive social media usage during the pandemic \cite{amitEarly2021}. Filipinos turned to online means that augmented with their regular routines for work, education, and shopping \cite{caparinoFilipinos2021}. Another non-pharmaceutical intervention that organically developed during the pandemic were practices of the public looking up information online \cite{galidoExploring2021}. There were significant interest at the onset of the pandemic, considering a majority of the information about COVID-19 and its societal implications were novel. Specifically, search engines such as Google has been proven to be a reliable public surveillance tool for virtual crowd behavior \cite{sofi-mahmudiAssessing2023}. It is worth acknolwedging that the vast majority of infodemiological studies probe selected \emph{individual} keywords on how popular they are on a search engine \cite{zayedGoogle2023}.

The aim of this investigation is to determine whether the network statistics of Google Trends (GT) keyword popularity scores could possibly match the temporal evolution of COVID-19 disease cases through the pattern recognition technique called dynamic time warping (DTW). The additional Sakoe-Chiba radius restriction offers a computationally inexpensive attribute to DTW because it only searches within the temporal space it was given. Another contribution of this study is that it pinpoints a specific geographic location in a country, which is in Metro Manila (an administrative region that has 16 cities and one municipality) with a total population of at least 13.4 million \cite{philippinestatisticsauthorityHighlights2021}. This deviates from the common scope used in the literature because much of them focused more on entire countries \cite{saegnerForecasting2022}. Through the best of our knowledge, this is the first opportunity to compare retroactive COVID-19 cases during the pandemic and networked online search behavior in the Philippines through a \emph{naive} computation such as DTW.

\section{Related Literature}

\subsection{Google Trends Mechanism and Reliability}

Google Trends (GT) is made available for anyone to find a search term's relative popularity over a certain place and time \cite{rovettaGoogle2024}. Any user across the world may download a specific time-series data of popularity scores based on the keywords or topics that the user queried \cite{behnenExperimental2020}. These series of popularity scores are called \emph{relative search volume} (RSV) integers that are bounded inclusively within 0 to 100 \cite{cebrianAddressing2024}. There are several attributes that the user may customize to enhance retrieving the GT data: place (this could be filtered from worldwide until down to the city level), query category (such as the parent categories of ``Health'', ``Finance'', and ``Arts \& Entertainment'' among others), time period (such that the user may input its preferred start and end dates), and the type of search (limited to ``Web Search'', ``Image Search'', ``News Search'', ``YouTube Search'', or ``Google Shopping'') \cite{alibudbudGoogle2023}. 

For the GT query categories, there are an estimated 1,400 possible combinations to choose from because each parent category has at most four hierarchical levels deep \cite{lolicDIY2024}. For example, someone may go to ``Business \& Industrial'' \textgreater \, ``Computers \& Electronics'' \textgreater \, ``Networking'' \textgreater \, ``VPN \& Remote Access'' as one way to filter out the popularity of a search term under that context. GT data is also \emph{anonymized}, which it will then process and aggregate the data in samples instead of the entire population \cite{cebrianAddressing2024}. It is said that Google receives billions of search requests worldwide and at any given moment. Therefore, it is computationally more feasible to calculate a subset of the search inputs at that specific time span \cite{lolicDIY2024}. Additionally, the GT data may return at different time intervals depending on the time period the user imposed. These are in hourly, daily, weekly, and monthly temporal configurations \cite{cebrianAddressing2024}. If someone retrieves a time period only from the previous seven days, then the hourly RSV score intervals are returned \cite{alibudbudGoogle2023}. There is a maximum of nine months of GT data for someone to download the daily RSV intervals. Then beyond nine months, the weekly data is returned. And lastly, monthly GT data is used for timelines beyond 5.25 years \cite{eichenauerObtaining2022}. 

Conflating these two points, there is a high chance of a potential problem to receive inconsistent GT data. The random sampling algorithm from Google is proprietary, which it is prone to sample variations and may affect the results \cite{eichenauerObtaining2022}. Besides this, the calculation of the RSV frequencies regardless of time formats have different data resolutions against one another \cite{lolicDIY2024}. Either way, data preprocessing techniques such as normalization are reliable ways to overcome these obstacles and preserve the data \cite{cebrianAddressing2024}. Thus, GT is still a frontier tool to assist the investigation and forecasting of a potential disease outbreak in an area \cite{rovettaGoogle2024}. Google represents 84.08\% of the global market share of the search engine industry  \cite{alibudbudGoogle2023}. This is its biggest advantage because it provides an almost instantaneous real-time statistics to supply initiatives that predict social phenomena \cite{lolicDIY2024}. It reliably estimates public perceptions, interests, and behaviors which are helpful for decision-based processes \cite{medeirosProper2021}. 

\subsection{Dynamic Time Warping in Health Analytics}
Various initiatives pursued DTW either as the primary tool for disease case surveillance or as an additional tool for a different purpose. Tracking human norovirus, which is the leading cause of acute gastroenteritis in the United States, was the aim of a particular study in Detroit, Michigan \cite{guzmanTracking2024}. The proponents applied dynamic time warping (DTW) analysis to compare waste water concentrations against Google search trends, and clinical cases whether if such relationship exists. DTW discovered that norovirus levels in wastewater, after being normalized by fecal indicators, produced the most similar trajectories to both the region’s positive cases data and internet searches for ``norovirus''. 

In terms of mental health, depression symptom dynamics were focused on a cohort of psychiatric inpatients \cite{hebbrechtUnderstanding2020}. DTW was applied to account for possible mismatch in symptom onset or improvement. The computed DTW distance for each symptom pair measured how similar the two symptoms' severity changed across repeated assessments. It unveiled which sets of symptoms tended to peak or subside in unison within and across patients. As a consequence, this approach facilitated more personalized clinical insights about their core depressive symptoms.

On the other hand, there were instances of specific DTW applications on COVID-19 incidences worldwide. The public interest in COVID-19 (gauged through Google Trends data) were lined up with national-level COVID-19 case and mortality data in six Western nations \cite{ziakasPublic2023}. DTW was implemented to check the shape-based similarity between the time series of Google search popularity and the time series of COVID-19 deaths and cases, and the government stringency index. They learned that public interest (Google searches) was most closely aligned with mortality. Then, public interest was less closely aligned with raw incident cases, and even lesser against the stringency index. 

In Poland, there was an enactment of DTW that focused on the 16 provinces of the country. It analyzed two time series in each voivodeship (province): the cumulative number of infected individuals, and the cumulative number of deaths \cite{landmesserUse2021}. Thanks to the DTW distances, it showed that the regions can be grouped based on the shape (and timing) of case and death curves, indicating that local characteristics (such as population density, and proximity to large cities) likely influence how quickly or severely the virus spreads.

Another unique implementation is using DTW as a strategy to cluster COVID-19 cases among 50 nations \cite{miralles-pechuanForecasting2023}. Here, the aim of DTW is to identify the set of countries whose daily-cases curves most closely resembled the country to be predicted. By clustering countries in ``similar epidemic shapes,'' it was hypothesized that the relevant patterns would transfer better to the target country’s forecast. The ramifications of the ``time-series filtering'' approach (using DTW to pick the most-similar countries for training) significantly boosted forecasting accuracy. Depending on the machine-learning method, RMSE was reduced by roughly 74\% once the DTW-based selection of similar countries were incorporated.

\section{Methodology}

\subsection{Google Trends Data Retrieval and Preprocessing}

The \texttt{pytrends} Python (version 3.12) library \cite{Pytrends} was imported to automatically scrape the relative search volume (RSV) values of all 15 keywords. These search terms were deliberately chosen so that it satisfies five categories: ``cough'', ``fever'', ``flu'', ``headache'', and ``rashes'' for the \textbf{English Symptoms} classification. The second category is its \textbf{Filipino Symptoms} counterpart, which consists of ``lagnat'' (fever), ``sipon'' (runny nose), and ``ubo'' (cough). For terms on \textbf{Face Wearing}, it includes ``masks'' and ``face shield''. The \textbf{Quarantine} category has only two search terms: ``ecq'' (Enhanced Community Quarantine) and ``quarantine''. Lastly, the \textbf{New Normal} category monitored regular life implications of the pandemic such as ``frontliners'', ``social distancing'', and ``new normal''. Each search term has a daily RSV time-series data from March 16, 2020 until March 15, 2021 depending on what data preprocessing is used. All the scraping occurred on a 30-day rolling window wherein each comma separated value (CSV) file were segmented beginning on each day within the one-year timeframe until this 30-day window data retrieval is not possible anymore.

Two data preprocessing techniques were used to address the Google Trends' inherent limitation on temporal resolution. First is the \emph{Rescaling Daily Data Method}, which was derived from \cite{brodeurCOVID192021} when they were ``stitching'' together different sets of weeks to form a cohesive timeline. The method name came to be because it leverages the relationship between daily and weekly GT data to fulfill calibration factors. The weekly GT data, which spans the entire study period in a single query, provides a consistent scaling reference against which the segmented daily data can be calibrated. This plan yielded multiple CSV files per keyword, each containing daily RSV data for a discrete 30-day segment, collectively spanning the entire study period. Weekly RSV data was also gathered directly from GT by requesting the entire period (March 16, 2020, to March 15, 2021) in a single query.

The second data preprocessing strategy adopted was the \emph{Merged Search Volume} (MSV) algorithm \cite{chuEnhancing2023}. The inventors of the MSV presented a new way to preserve the daily data (or any format) for an extended length of time. Unlike the Rescaling Daily Data Method, which calibrates daily data using weekly references, the MSV transformation \emph{directly addresses} the temporal inconsistency by estimating correction factors between consecutive data segments. This is helpful when seeking to maintain the relative proportionality of daily fluctuations.


\subsection{Conversion of Google Trends Data into Network Statistics}

Following the preprocessing of Google Trends data using both transformations, dynamic network graphs were constructed to model the temporal interconnectedness among COVID-19-related search terms. This method extends beyond traditional time-series analysis because it incorporates the complex interdependencies between multiple search queries simultaneously \cite{soTopological2021}. To capture the evolving relationships between queries, a rolling-window was implemented to construct the time-dependent networks. For each time point $t$, 15-day and 30-day retrospective correlation windows of preprocessed daily data were considered for all 15 keywords. Using the preprocessed daily RSV data, the distance correlation matrix $\rho_t$ was computed between all pairs of keywords at each time point $t$ using the \texttt{dcor} Python package (version 3.12) \cite{ramos-carrenoDcor2023}. 


To convert these correlation measurements into a network structure, there was a minimum requirement of reaching at least the threshold to establish edges between keyword nodes. The adjacency matrix $A_{i,j,t}$ among search terms $i$ and $j$ at date $t$ was implemented such that if the correlation value is greater than or equal to $\theta$ (correlation threshold), then it merits an undirected edge between $i$ and $j$ for that graph \cite{chuEnhancing2023}. To comprehensively investigate the sensitivity of network metrics to varying levels of connection strength, there was a systematic computation of threshold values: $\theta \in \{0.4, 0.5, 0.6, 0.8\}$. The multi-threshold parameters enable assessments of network dynamics across different stringency levels \cite{soTopological2021}. Lower thresholds ($\theta = 0.4$) capture weaker associations, resulting in denser networks that may include spurious connections. Higher thresholds ($\theta = 0.8$) impose more restrictive criteria, as it yields sparser networks with only the strongest associations preserved. Intermediate thresholds ($\theta = 0.5, 0.6$) provide balanced perspectives, with $\theta = 0.5$ serving as the standard threshold in the network analysis literature \cite{soVisualizing2020}.

To quantify the macroscopic properties of these networks, two complementary network statistics were calculated: \emph{network density} and \emph{clustering coefficient}. The metrics below provide meaningful insights into its structural characteristics. Network density ($D_t$) represents the proportion of potential connections that are actually realized in the network at time $t$ \cite{bedruBig2020}:

\begin{equation}
D_t = \frac{2E_t}{N(N-1)}
\end{equation}

where $E_t$ is the number of edges in the network at time $t$, and $N(N-1)/2$ represents the maximum possible number of edges in an undirected network with $N$ nodes. The values range from 0 (no connections) to 1 (fully connected network). This metric quantifies the overall cohesiveness of health-seeking behavior across the multiple coronavirus pandemic keywords. Periods of high network density suggest synchronized interest in multiple related search terms, potentially indicating collective public response to significant pandemic developments \cite{chuDetecting2020}. While network density characterizes overall connectedness, the clustering coefficient ($C_t$) provides insight into the network's local structure by quantifying the tendency of nodes to form tightly connected groups or clusters \cite{wattsCollective1998}. For each node $v$, there exists $\lambda(v)$, the number of triangles formed with that node, and $\tau(G)$, the total number of triplets across the entire graph. The global clustering coefficient is then calculated as \cite{chalanconClustering2013}:

\begin{equation}
C_{t}(G) = \frac{\sum_{v\in V}\lambda(v)}{\tau(G)} = \frac{1}{V}\sum_{v\in V}c(v)
\end{equation}

The clustering coefficient aggregates these local measures, weighted by each node's connectivity, to provide a network-level assessment of clustering. High values indicate that search terms tend to form cohesive groups, suggesting thematic clusters in health-seeking behavior. Low values suggest more dispersed or random patterns of association between search terms \cite{soTopological2021}. In the context of the online searches, high clustering may indicate distinct thematic groups of search terms (e.g., symptoms-related queries clustering separately from policy-related terms), while lower clustering despite high density would suggest more uniform associations across different categories of the keywords.

\subsection{Retrieving COVID-19 Daily Confirmed and Active Cases}

The Department of Health (DOH) owns a public repository of COVID-19 cases called the data drop, wherein its epidemiology bureau (EB) collects the number of patients who got infected from hospitals nationwide and releases them at noon \cite{departmentofhealthoftherepublicofthephilippinesPrivacy}. This initiative ended in 2023, however the repository is still available at the time of writing \cite{departmentofhealthoftherepublicofthephilippinesCOVID19}. The Metro Manila data was extracted from here such that the \texttt{RegionRes} (``region of residence'') and \texttt{ProvinceRes} (``province of residence'') must be equal to \texttt{NCR}. The \texttt{DateRepConf} (``date of report confirmation'') column is the most crucial variable for determining the daily \emph{confirmed} cases because it tells the date in which the patient has been publicly announced as a positive case of the coronavirus \cite{galarosaHow2020}. Opposite to the reported infection of an individual is the \texttt{DateRepRem} (``date of report of removal'') variable, which tells the instance of when was the recovery or death of that patient happened \cite{upmediaandpublicrelationsofficePrevailing2020}. Both \texttt{DateRepConf} and \texttt{DateRepRem} are important to calculate the \emph{active} coronavirus cases as the pandemic progressed.

Daily confirmed cases represent the incidence of newly diagnosed COVID-19 cases on each specific date. This metric reflects the rate at which new infections are being detected within the population. In contrast, active cases represent the prevalent burden of COVID-19 at any given time point. It is the accumulated number of confirmed cases minus those who have been removed from the active case pool through recovery or mortality. This metric provides a snapshot of the concurrent disease burden within the community. For each date in this sequence, two key parameters were quantified: the number of new cases confirmed on that date and the number of cases removed (through recovery or death) on that same date. The active case count was then calculated through a cumulative approach, where $A_t = A_{t-1} + C_t - R_t$, such that $A_t$ represents the active cases on day $t$, $C_t$ denotes new confirmed cases on day $t$, and $R_t$ signifies removed cases on that same day $t$. This recurrence relation was initialized at zero and computed iteratively across the entire time series, producing a dynamic representation during the pandemic.

\subsection{Implementation of Dynamic Time Warping}

Dynamic time warping (DTW) is commonly used to quantify similarities between time series of different lengths that may exhibit temporal distortions \cite{seninDynamic}. Consider two time series $X = (x_1, x_2, \ldots, x_N) \in \mathbb{R}^N$ of length $N$ representing daily disease case (active or confirmed) counts and $Y = (y_1, y_2, \ldots, y_M) \in \mathbb{R}^M$ of length $M$ denoting Google Trends data. DTW aims to find an optimal alignment between these sequences by warping the time axis. Formally, DTW identifies a warping path $P = (p_1, p_2, \ldots, p_L)$ where each element $p_l = (i_l, j_l) \in [1:N] \times [1:M]$ pairs index $i_l$ from sequence $X$ with index $j_l$ from sequence $Y$ \cite{mullerDynamic2007}. A valid warping path must satisfy the following constraints:

\begin{itemize}
   \item \textbf{Boundary condition}: $p_1 = (1,1)$ and $p_L = (N,M)$, ensuring the alignment considers complete sequences.
   
   \item \textbf{Monotonicity condition}: $i_1 \leq i_2 \leq \ldots \leq i_L$ and $j_1 \leq j_2 \leq \ldots \leq j_L$, preserving the time-ordering of points.
   
   \item \textbf{Continuity condition}: $p_{l+1} - p_l \in \{(1,0), (0,1), (1,1)\}$ for $l \in [1:L-1]$, limiting the warping to adjacent cells.
\end{itemize}

The DTW distance between $X$ and $Y$ is then defined as the minimum cumulative distance over all possible warping paths \cite{landmesserUse2021}:

\begin{equation}
\text{DTW}(X,Y) = \min_{P \in \mathcal{P}} \sum_{l=1}^{L} d(x_{i_l}, y_{j_l})
\end{equation}

where $\mathcal{P}$ is the set of all valid warping paths and $d(x_{i_l}, y_{j_l})$ represents the distance between elements $x_{i_l}$ and $y_{j_l}$. Given the inherent differences in scale between these series, the disease case data requires normalization prior to comparison. The min-max normalization was applied to the disease case time series as follows:

\begin{equation}
X_{\text{norm}} = \frac{X - \min(X)}{\max(X) - \min(X)}
\end{equation}

The Google Trends data is maintained in its original form as it is because the network statistics values are already between 0 and 1. It is known that DTW employs dynamic programming to efficiently compute the optimal warping path \cite{seninDynamic}. The approach constructs an accumulated cost matrix $D \in \mathbb{R}^{N \times M}$ where each element $D(i,j)$ represents the minimum cumulative distance for aligning subsequences $X_{\text{norm}}(1:i)$ and $Y(1:j)$. The local cost matrix $C \in \mathbb{R}^{N \times M}$ was first computed:

\begin{equation}
C(i,j) = |x_{\text{norm},i} - y_j|, \quad i \in [1:N], j \in [1:M]
\end{equation}

Then, the accumulated cost matrix is populated recursively:

\begin{equation}
D(i,j) = C(i,j) + \min \begin{cases}
D(i-1,j)\\
D(i,j-1)\\
D(i-1,j-1)
\end{cases}
\end{equation}

with base case $D(1,1) = C(1,1)$. Typically, unrestricted DTW is oftenly practiced. However, applying this characteristic produces alignments where a single point in one series maps to a large subsection of the other. To prevent this and improve computational efficiency, the Sakoe-Chiba band constraint was considered\cite{sakoeDynamic1978}, which restricts the warping path to a band along the main diagonal $|i - j| \leq r$, wherein $r$ is the radius parameter defining the maximum allowable temporal deviation in terms of the number of days. This constraint modifies the accumulated cost matrix calculation:

\begin{equation}
D(i,j) = \begin{cases}
C(i,j) + \min\{D(i-1,j),\\
\quad D(i,j-1), D(i-1,j-1)\}, & \text{if } |i-j| \leq r\\
\infty, & \text{if } |i-j| > r
\end{cases}
\end{equation}

For our analysis of the relationship between disease cases and Google network statistics behavior, multiple radius values were evaluated: $r \in \{7, 15, 20, 30, 50\}$ days. The constraint creates a corridor of width $2r+1$ along the matrix diagonal, significantly reducing the search space from $O(NM)$ to $O(Nr)$ wherein $r < M$. After computing the accumulated cost matrix, the optimal warping path is retrieved through backtracking. The resulting path $P^* = (p_1, p_2, \ldots, p_L)$ provides the optimal element-wise mapping between disease case patterns and networked search behavior indicators. This implementation uses Python with the \texttt{numpy} package for matrix operations. The Sakoe-Chiba constraint is applied via a Boolean mask matrix: 

\begin{equation}
M(i,j) = \begin{cases}
\texttt{False}, & \text{if } |i-j| \leq r\\
\texttt{True}, & \text{if } |i-j| > r
\end{cases}
\end{equation}

where cells marked \texttt{True} are excluded from the path calculation. For each disease-search term pair, DTW distances were determined across all radius values to determine the optimal temporal constraint. Whatever the result of the distance is, this serves as the similarity metric. It tells that lower values have stronger correspondence between disease patterns and search behavior. Provided below is Algorithm \ref{alg:dtw_sakoe_chiba} that summarizes how the DTW was evaluated between the two time-series.

\begin{algorithm}[h!]
\caption{DTW with Sakoe-Chiba Band}
\label{alg:dtw_sakoe_chiba}
\begin{algorithmic}[1]
\REQUIRE Series $X \in \mathbb{R}^N$, $Y \in \mathbb{R}^M$, radius $r$
\ENSURE DTW distance $D(N,M)$ and optimal path $P^*$
\STATE $X_{\text{norm}} \gets \frac{X - \min(X)}{\max(X) - \min(X)}$
\STATE Initialize cost matrix $C(i,j) \gets |x_{\text{norm},i} - y_j|$ for all $i,j$
\STATE Initialize $D$ with $\infty$; $D(1,1) \gets C(1,1)$
\FOR{$i=1$ to $N$, $j=1$ to $M$ where $|i-j| \leq r$}
    \IF{$(i,j) \neq (1,1)$}
        \STATE $\!\begin{aligned}
        D(i,j) \gets C(i,j) + \min\begin{cases}
        D(i-1,j)   & \text{if } i>1 \\
        D(i,j-1)   & \text{if } j>1 \\
        D(i-1,j-1) & \text{if } i,j>1
        \end{cases}
        \end{aligned}$
    \ENDIF
\ENDFOR
\STATE Backtrack from $(N,M)$ to $(1,1)$ to build $P^*$ 
\RETURN $D(N,M)$, $P^*$
\end{algorithmic}
\end{algorithm}

\section{Results and Discussion}

The network metrics were constructed from Google Trends data using both the Rescaling Daily Data Method and Merged Search Volume (MSV) algorithm transformations. We systematically explored 320 distinct parameter configurations to identify optimal settings for epidemic surveillance through network analysis. The parameter space encompassed two network metrics (network density and clustering coefficient), two data source transformations (Rescaling Daily Data and MSV), four threshold values (0.4, 0.5, 0.6, and 0.8), two correlation window sizes (15 and 30 days), two case comparison types (Confirmed Cases and Active Cases), and five Sakoe-Chiba band radius values (7, 15, 20, 30, and 50 days). DTW scores in the dataset ranged from 36.30 to 101.67. The mean DTW score across all configurations was 62.30. This mean serves as the baseline for assessing relative performance.

\subsection{Statistical Significance of the Parameters}

To systematically evaluate the impact of each parameter on DTW performance, the non-parametric Kruskal-Wallis tests were conducted, which serves as an alternative to one-way ANOVA \cite{hoffmanChapter2019}. Table \ref{tab:kruskal_results} presents the results, revealing that five of the six parameters significantly influenced alignment quality ($p < 0.05$).

The most influential parameter was the disease case comparison dimension ($H = 154.80$, $p = 1.55 \times 10^{-35}$), with confirmed cases yielding substantially better alignment than active cases (average DTW scores of 51.08 and 73.53, respectively). This finding suggests that network metrics derived from Google Trends activity in Metro Manila correlate more closely with new case identification than with the prevalence of active infections. This pattern may reflect public health-seeking behavior in response to official case announcements, which typically emphasize daily new cases rather than active case counts in health-seeking behavior. 

Simultaneously, the threshold values parameter significantly impacted performance ($H = 27.71$, $p = 4.17 \times 10^{-6}$), with lower thresholds of 0.5 (avg. DTW = 58.30) and 0.4 (avg. DTW = 58.48) generally outperforming more restrictive thresholds of 0.8 (avg. DTW = 71.11). This indicates that moderate thresholds for establishing connections between search terms in our network model captured more meaningful dynamics during the pandemic in Metro Manila. Too high thresholds may have eliminated important but moderate correlations between search terms, while moderate thresholds preserved these relationships.

\begin{table}[!t]
\caption{Kruskal-Wallis Test Results for Parameter Significance}
\label{tab:kruskal_results}
\centering
\begin{tabular}{lrrr}
\hline
\textbf{Variable} & \textbf{$H$-statistic} & \textbf{$p$-value} & \textbf{Significant?} \\
\hline
Network Statistic Type & 26.44 & 2.72e-7 & Yes \\
Data Preprocessing Technique & 4.10 & 4.30e-2 & Yes \\
Threshold & 27.71 & 4.17e-6 & Yes \\
Correlation Window & 1.66 & 1.98e-1 & No \\
Disease Case Comparison & 154.80 & 1.55e-35 & Yes \\
Sakoe-Chiba Radius & 20.97 & 3.21e-4 & Yes \\
\hline
\end{tabular}
\end{table}

The network statistic parameter emerged as another significant factor ($H = 26.44$, $p = 2.72 \times 10^{-7}$), with the network density (avg. DTW = 57.07) consistently outperforming against the clustering coefficient (avg. DTW = 67.54). This suggests that the overall density of the search query network was more predictive of epidemic progression in Metro Manila than the tendency of search terms to form tightly connected clusters. During a pandemic where information-seeking behavior spans multiple related topics, the connectivity between topics appears to be more informative than the clustering of specific keyword nodes.

The radius parameter for the Sakoe-Chiba band exhibited significant impact ($H = 20.97$, $p = 3.21 \times 10^{-4}$), with larger radii generally yielding better performance. The average DTW score decreased monotonically as radius increased: 67.75 for the 7-day limitation, 64.11 for the 15-day restriction, 62.62 for the 20-day attribute, 60.14 for the 30-day, and 56.90 for the 50-day imposed parameter. This trend indicates that broader temporal alignment constraints enhance the correspondence between network metrics and epidemic curves. It also likely accommodates more the lag between disease incidence and related online search behavior in Metro Manila's context. The optimal performance at the Sakoe-Chiba restriction of 50 days suggest that public health-seeking behavior has a complex temporal relationship with case trajectories that spans several weeks.

The method on what data preprocessing transformation used also significantly influenced performance ($H = 4.10$, $p = 4.30 \times 10^{-2}$), with the Rescaling Daily Data Method (avg. DTW = 60.48) providing better alignment than MSV (avg. DTW = 64.13). This suggests that the Rescaling Daily Data Method, which leverages the relationship between daily and weekly Google Trends data, better preserves the temporal dynamics relevant to epidemic surveillance than the MSV algorithm. On the other hand, the correlation window size parameter interestingly did not significantly affect performance ($H = 1.66$, $p = 1.98 \times 10^{-1}$), indicating that both 15-day and 30-day windows adequately capture the relevant temporal dynamics for alignment. This robustness to window size variation is advantageous for practical implementation so that it offers some flexibility in temporal aggregation.

\subsection{Optimal Parameter Configurations}

\begin{figure}[!h]
    \centering
    \includegraphics[width=3.5in]{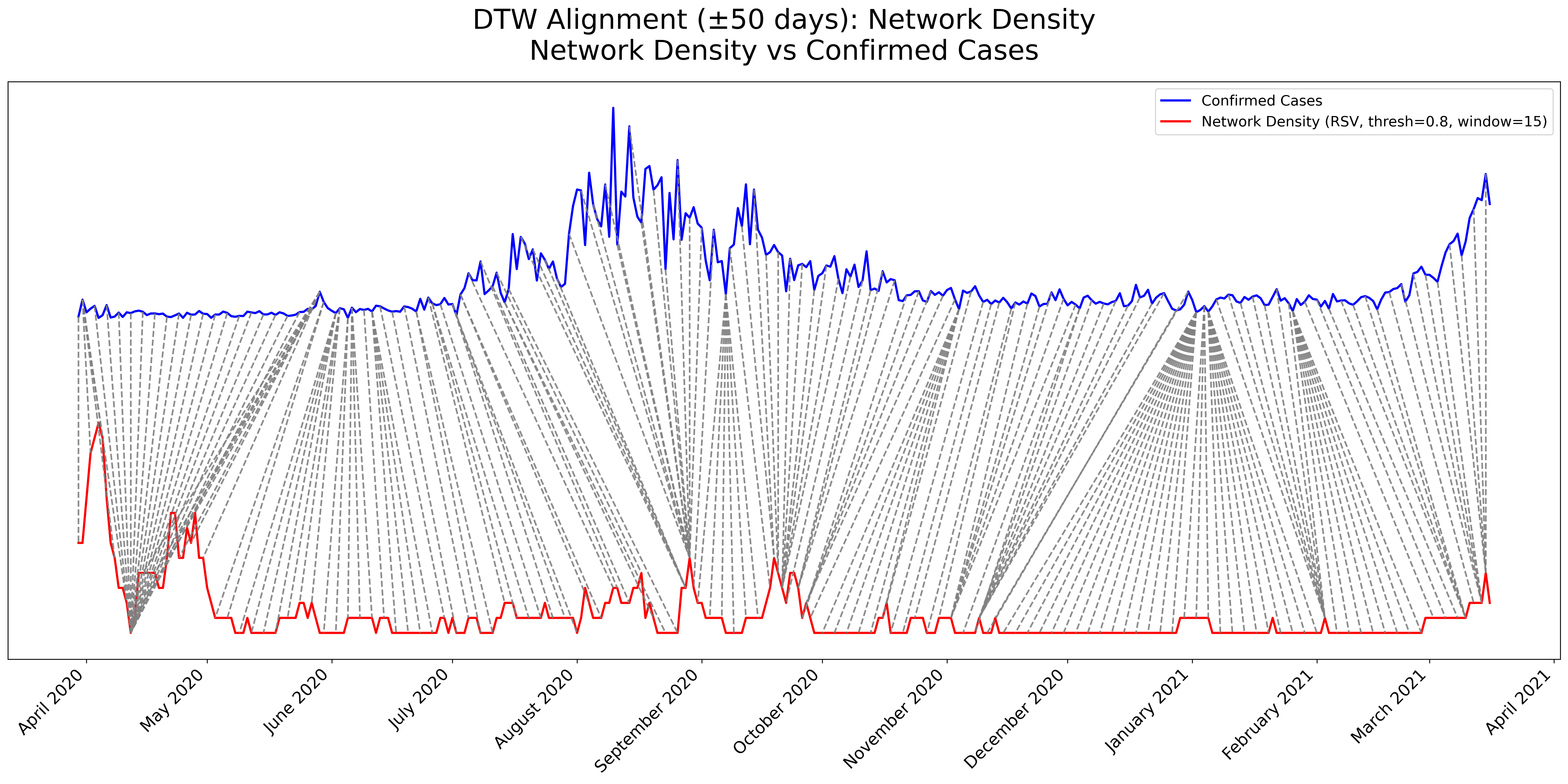}
    \caption{Temporal Alignment Between Network Density Values (Rescaling Daily Data Method, Threshold = 0.8, Correlation Window = 15 days, Sakoe-Chiba Radius = 50 days) with Confirmed COVID-19 Cases in Metro Manila from March 2020 to March 2021.}
    \label{fig:best_alignment_confirmed}
\end{figure}

\begin{figure}[!h]
    \centering
    \includegraphics[width=3.5in]{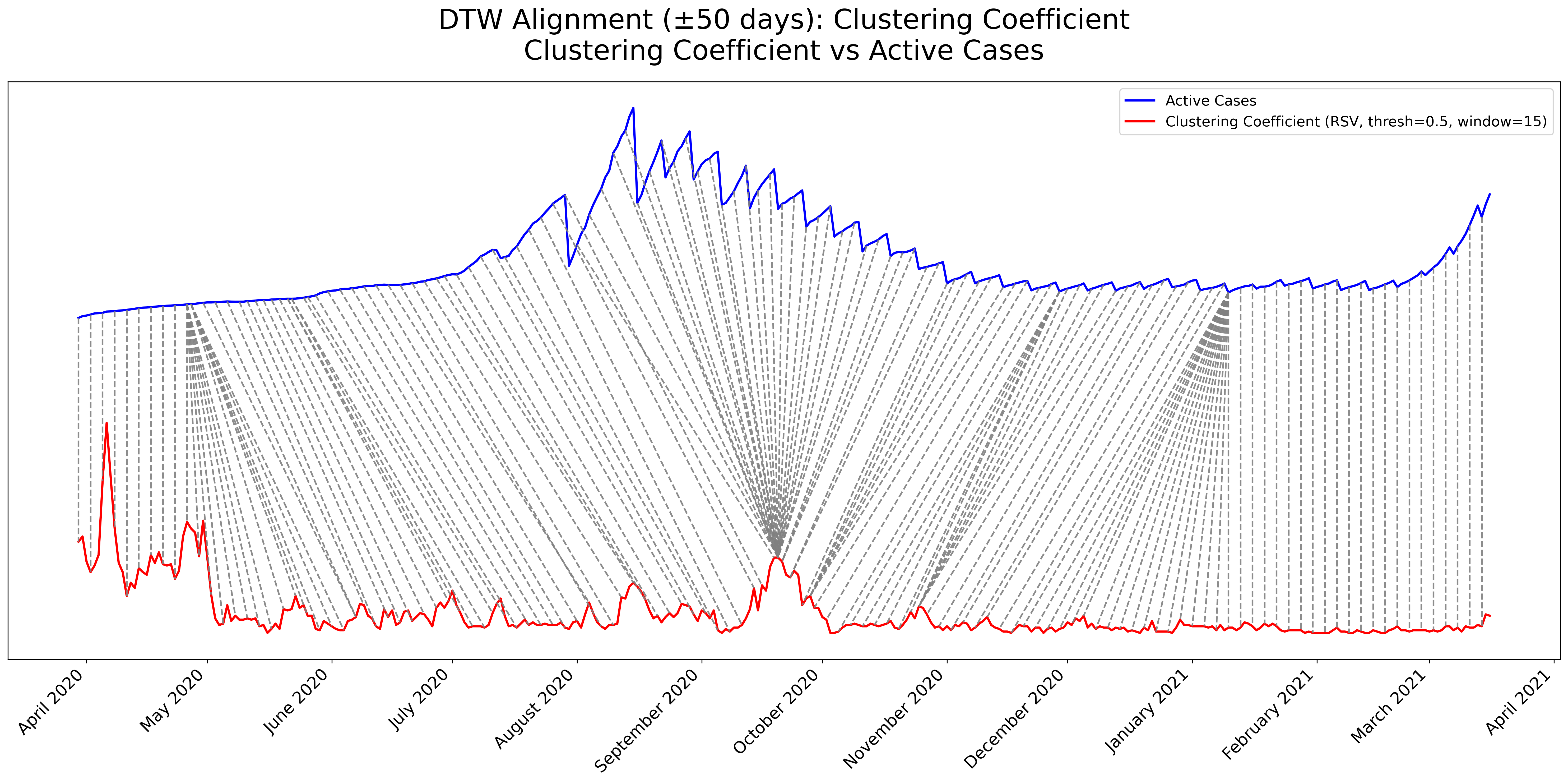}
    \caption{Temporal Alignment Between Clustering Coefficient Values (Rescaling Daily Data Method, Threshold = 0.5, Correlation Window = 15 days, Sakoe-Chiba Radius 50 day) with Confirmed COVID-19 Cases in Metro Manila from March 2020 to March 2021.}
    \label{fig:best_alignment_active}
\end{figure}

Table \ref{tab:optimal_configs} presents the optimal parameter configurations for each combination of metric type and comparison type. Across all combinations, several patterns emerge: larger radii (particularly 50) consistently yield better alignment, the Rescaling Daily Data outperforms MSV as a data transformation method, and optimal thresholds vary based on the specific metric-comparison combination.

\begin{table}[!h]
\caption{Optimal Parameter Configurations}
\label{tab:optimal_configs}
\centering
\begin{tabular}{@{}lccccc@{}}
\hline
\textbf{Configuration} & \textbf{Data Preprocess} & \textbf{Thr} & \textbf{Win} & \textbf{Rad} & \textbf{DTW} \\
\hline
ND - Confirmed & Rescaling Daily & 0.8 & 15 & 50 & 36.30 \\
ND - Active & Rescaling Daily & 0.6 & 15 & 50 & 42.68 \\
CC - Confirmed & Rescaling Daily & 0.5 & 15 & 50 & 36.46 \\
CC - Active & Rescaling Daily & 0.4 & 15 & 50 & 51.88 \\
\hline
\multicolumn{6}{@{}l@{}}{\footnotesize ND: Network Density, CC: Clustering Coefficient} \\
\multicolumn{6}{@{}l@{}}{\footnotesize Src: Source, Thr: Threshold, Win: Correlation Window, Rad: Radius} \\
\end{tabular}
\end{table}

The best overall performance (DTW = 36.30) was achieved with network density, using the Rescaling Daily Data transformation, a threshold of 0.8, a 15-day correlation window, and a radius of 50 days when aligning with confirmed cases. This configuration represents the most promising approach for operationalizing network metrics as epidemic surveillance tools in Metro Manila. Figure \ref{fig:best_alignment_confirmed} illustrates the temporal alignment between this network metric configuration and the confirmed COVID-19 case trajectory.

For active cases, which generally proved more challenging to align with network metrics, the optimal configuration utilized network density from the Rescaling Daily Data Preprocessing Method with a lower threshold of 0.6, 15-day correlation window, and a 50-day Sakoe-Chiba radius (DTW = 42.68). The need for a lower threshold when aligning with active cases may reflect the differing nature of online searches. Meanwhile, clustering coefficient metrics exhibited their best performance when utilizing the same Rescaling preprocessing method with moderate thresholds (0.5 for confirmed cases, 0.4 for active cases), consistently favoring a 15-day correlation window and 50-day Sakoe-Chiba radius. The optimal DTW scores for the clustering coefficient statistic (36.46 for confirmed cases, 51.88 for active cases) were comparable to or slightly worse than those achieved with the network density. This again reinforces the preference for density-based metrics in this application.

\subsection{Implications for Epidemic Surveillance in the Philippine Context}

The results demonstrate that Google Trends-based network metrics, when optimally configured, can achieve substantial temporal alignment with COVID-19 case trajectories in Metro Manila. The best configurations achieved DTW scores in the 36 to 52 range, representing significantly better alignment than would be expected by chance. It suggests that network metrics derived from online search behavior data can serve as complementary indicators for traditional epidemic surveillance systems in the Philippine setting. These results are particularly relevant given the Philippines' extensive digital presence during the pandemic \cite{amitEarly2021}. As Filipinos increasingly turned to online means to support their regular routines for health-seeking actions during strict community quarantine measures \cite{caparinoFilipinos2021}, the digital traces left became increasingly valuable for public health monitoring.

The superior performance of the network density metrics emphasizes the value of overall connectivity patterns in tracking epidemic progression. As information about disease outbreaks diffuses throughout online, the density of connections appears to connect strongly with case trajectories. Thus, it has the potential of serving as a leading indicator for epidemic waves. This insight implies that changes in online health-seeking patterns may often respectively precede official case reporting.

\section{Conclusion}

This comprehensive parameter exploration for aligning Google Trends search popularity network metrics with COVID-19 case trajectories in Metro Manila provides key insights for epidemic surveillance in a local context. First, the network density consistently outperforms the clustering coefficient for this application, though both metrics can achieve strong alignment when optimally configured. Second, confirmed cases generally align better with network metrics than active cases, suggesting that online health-seeking activity more closely tracks new case identification than ongoing prevalence. Third, larger temporal alignment radii (such as 50 days) consistently improve mapping quality. It has repercussions on how anticipating future disease cases could take a significant number of days prior. In return, government health agencies and officials may benefit on the proactivity of detecting cases way ahead of time.

The significance of five out of six parameters in the analysis underscores the importance of careful parameter selection when implementing network-based surveillances. The non-significance of window size provides valuable flexibility for practical implementation, allowing system designers to balance temporal resolution against computational efficiency without compromising alignment quality. Therefore, our results demonstrate that a network metric approach towards Google Trends data, when appropriately configured, can serve as valuable complementary indicators for epidemic surveillance in the Metro Manila region. By capturing the complex patterns of public information-seeking behavior during a health crisis, these metrics provide insights beyond traditional surveillance data, which has the potential to discover early warning capabilities and appropriately strategize on its communication strategies for future outbreaks.

\section*{Acknowledgments}


The authors forward its gratitude to the Ateneo Social Computing Science Laboratory and its parent organization, the Ateneo Center for Computing Competency and Research, for the opportunity and platform of spearheading this study. Much of the appreciations are also extended to Christian Pulmano and Atty. Kennedy Espina for all their suggestions and recommendations shared towards the success of this endeavor.

\bibliographystyle{IEEEtran}
\bibliography{references}

\end{document}